\documentclass[a4paper,11pt]{article}
\usepackage{pos}
\usepackage{graphicx}	
\usepackage{amsmath}	
\usepackage{natbib} 

\usepackage{comment}
\usepackage{soul}
\usepackage{graphicx}

\title{Dissecting the radiation mechanism of short GRB~160821B through multi-wavelength modelling}
\ShortTitle{Multi-wavelength modelling of GRB~160821B}

\author*[a]{Ankur Ghosh}
\author[a,b]{Monica Barnard}
\author[c]{Jagdish C. Joshi}
\author[a,d,e]{Soebur Razzaque}

\affiliation[a]{Centre for Astro-Particle Physics (CAPP), Department of Physics,\\
PO Box 524, Auckland Park 2006, Johannesburg, South Africa}

\affiliation[b]{Centre for Space Research, North-West University, Potchefstroom 2520, South Africa}

\affiliation[c]{Aryabhatta Research Institute of Observational Sciences, Manora Peak, Nainital 263129, India}

\affiliation[d]{Department of Physics, The George Washington University,\\
Washington, DC 20052, USA}

\affiliation[e]{National Institute for Theoretical and Computational Sciences (NITheCS), South Africa}

\emailAdd{ghosh.ankur1994@gmail.com}
\emailAdd{srazzaque@uj.ac.za}

\abstract{GRB~160821B is the only short GRB detected to date at very high energy (VHE, $\gtrsim 100$ GeV). At a redshift $z=0.161$, it was detected by MAGIC telescopes 
approximately four hours since the trigger. VHE dataset was complied with the datasets of other wavelengths in between the timescale of 1.7 to 4 hours to construct the broadband spectral energy distribution (SED). In previous studies of GRB~160821B, synchrotron and external Compton (EC) model could explain the VHE emission better than the synchrotron and synchrotron self-Compton (SSC) model. Although, these fits were mostly eyeballing data without any optimisation. Our model includes the combination of synchrotron, SSC, and EC models with Markov Chain Monte Carlo (MCMC) techniques. Our analysis reveals that the EC contribution is negligible in comparison with the SSC and our model explains the VHE data well for the wind medium. We found that GRB~160821B is the least energetic VHE GRB and it occurred in high density wind medium which is quiet unusual for a short GRB. But like other long-duration VHE GRBs, GRB~160821B occurred in a poorly magnetised medium. As there is no statistical study on afterglow modelling of short GRB sample, we compare the inferred properties of GRB~160821B with other VHE GRBs. It stands out distinctively in the $E_{k, \rm iso}$ - $\epsilon_B$ parameter space and lies outside the 3-$\sigma$ region of the correlation. In future, more VHE detections of short GRBs, in the CTA era, will provide crucial insights into the emission sites, radiation mechanisms, and particle acceleration, as well as their connection to long GRBs.
}

\ConferenceLogo{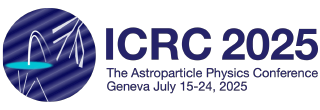}

\FullConference{39th International Cosmic Ray Conference (ICRC2025)\\
 15–24 July 2025\\
Geneva, Switzerland\\}

\begin{document}
\maketitle

\section{Introduction}
\label{sec1:intro}

The very high energy (VHE, $\gtrsim$ 100 GeV) $\gamma$-rays represent the most energetic part of the electromagnetic spectrum which play a critical role in the era of multi-wavelength and multi-messenger astrophysics. VHE $\gamma$-rays are important to investigate the particle acceleration and radiation mechanisms associated with $\gamma$-ray bursts (GRBs). These VHE photons are predominantly found during the afterglow phase when the jet interacts with the surrounding medium, although their exact origin is still not confirmed \citep{2022Galax..10...66M, 2022Galax..10...74G, 2022Galax..10...67B}.

So far, VHE photons have been detected in six long GRBs namely GRB~180720B, GRB~190114C, GRB~190829A, GRB~201015A, GRB~201216C, and GRB~221009A using the ground-based Cherenkov telescopes, such as High Energy Stereoscopic System (H.E.S.S.), Major Atmospheric Gamma Imaging Cherenkov Telescopes (MAGIC), and Large High Altitude Air Shower Observatory (LHAASO) \citep{Abdalla2019, 2019Natur.575..455M, HESS:2021dbz, doi:10.1126/sciadv.adj2778, 2020GCN.28659....1B, Abe2024}. Prior to the confirmed VHE detection from GRB~190114C, MAGIC searched for the VHE photons from three short GRBs (GRB~061217, GRB~100815A and GRB~160821B) at redshift $z<$1. But GRB~160821B ($z$ = 0.16) is the only short GRB for which VHE emission has been reported $\sim$ 4 hours after the explosion, at a $3\sigma$ confidence level \citep{Acciari2021ApJ..90A}. This raises an interesting open question: why do we not detect VHE photons from more nearby short GRBs?

Several radiation mechanisms have been proposed to interpret the VHE data, among which the synchrotron self-Compton (SSC) emission is mostly considered where synchrotron seed photons are up-scattered to higher energies \citep{2001ApJ...559..110Z, dermer_50GeV_IC, Panai2000ApJ543, sari_IC_paper, Nakar:2009er, 2019Natur.575..455M, Abdalla2019}. In some modelling, the external Compton (EC) emission is dominant that originates from the up-scattering of photons in external fields, e.g., cosmic microwave background (CMB), near-infrared (nIR) \citep{2010MNRAS.402L..54M, 2021ApJ...908L..36Z, zhang2021ApJ..55Z, 2023ApJ...947L..14Z, Barnard2024, 2025arXiv250518041B}. Along with that, the proton-synchrotron process has been explored in both the forward and reverse shock regions of GRB afterglows to explain the observed VHE emission \citep{2023ApJ...955...70I, Misra2021MNRAS_5685M, 2023ApJ...947L..14Z}. A photo hadronic mechanism involves interactions between VHE protons and ambient afterglow photons, leading to subsequent VHE radiation \citep{sahu2020ApJ.41S, 2022ApJ...929...70S, Klinger_2024}. However, such models usually demand energy budgets that surpass the total prompt $\gamma$-ray output observed from GRBs significantly, raising concerns about their viability.

To interpret the multi-wavelength spectral energy distributions (SEDs) of VHE detected long GRBs, a wide range of theoretical frameworks have been employed. These include both single-zone and multi-zone emission models, incorporating synchrotron, SSC, and EC processes, with different jet morphologies such as structured or two-component jets. Additionally, emission from forward and reverse shocks, as well as off-axis viewing geometries, have been investigated to interpret the complex afterglow behaviour observed in these events \citep{derishev2021ApJ.135D, Joshi2021, 2019Natur.575..455M, Salafia2022, Abdalla2019, Ren2023, Banerjee2025, 2025arXiv250518041B}. For the short GRB~160821B, \citet{Acciari2021ApJ..90A} employed a single-zone model involving synchrotron and SSC radiation to fit the broadband data. In contrast, \citet{2021ApJ...908L..36Z} proposed that the EC is a more viable mechanism to explain the VHE emission. These differing approaches reflect the diversity and complexity of VHE emission scenarios in short GRBs and emphasize the need for detailed multi-wavelength modelling to constrain the underlying physical processes.

In this work, we fit multiwavelength data from the population of VHE GRBs with synchro-Compton emission to the extent of a single-zone, namely the forward shock, scenario. For this purpose, we employ the publicly-available \textsc{naima} code~\citep{2015ICRC...34..922Z}~\footnote{Additional documentation on the code and the installation thereof is available at \url{http://naima.readthedocs.org}.} with some modifications. Our goal is to exhaustively explore this one-zone scenario by fitting as much broadband and multi-epoch data as possible for the VHE GRBs. This work, therefore, can identify cases where models with multi-emission zones are necessary. 

\section{Multi-wavelength modelling using NAIMA}
\label{sec2:NAIMA}

The model used in our study is the modified version of the afterglow emission model of GRB~190829A detected by H.E.S.S. \citep{HESS:2021dbz}. We applied the model to fit the multi-wavelength data of GRB~160821B to understand the particle acceleration, blast-wave evolution, and radiation mechanisms associated with short GRBs, adopting the formulations used in \citet{2025arXiv250518041B}. The code employs the \textsc{naima} package\footnote{\url{https://github.com/Carlor87/GRBmodelling}} to compute non-thermal emission, i.e., synchrotron and inverse Compton (IC) from a homogeneous population of relativistic electrons and protons. \textsc{Naima} uses the steady state single zone model with full Klein Nishina cross-section. 

Afterglow modelling is performed assuming emission originates from the forward shock of the blastwave after its deceleration time ($t_{\rm dec}$), where the system evolves following self-similar solutions in either a constant-density interstellar medium (ISM) or a wind-like environment with a density profile $\rho \propto R^{-2}$, following \citet{Blandford1976}. The evolution of $\Gamma_0$ and the deceleration radius for both mediums are described in Appendix A of \citet{2025arXiv250518041B}. As \textsc{naima} does not calculate the deceleration time ($t_{\rm dec}$), we use equation (1) and (2) of \citet{Joshi2021} to estimate $\Gamma_0$ for ISM and wind mediums. In the wind scenario, the normalization parameter is taken as $A = 3.02 \times 10^{35} A_*$ cm$^{-1}$, with $A_*$ values reported in Table~\ref{tab:model_parameters}, alongside the ISM densities ($n$) in cm$^{-3}$. 

The synchrotron radiation formalisms are adopted from \citet{2010PhRvD..82d3002A}, while the SSC and EC formalism with non-thermal and thermal photons are taken from \citet{1981ApnSS..79..321A} and \citet{2014ApJ...783..100K}. The detailed synchrotron, SSC, and EC models are explained in Appendix A3, A4, and A5 of \citet{2025arXiv250518041B}. The modifications introduced to compute the maximum electron energy are detailed in Appendix A6 of \citet{2025arXiv250518041B}. To account for high-energy attenuation, internal $\gamma\gamma$ absorption is calculated using the methods of \citet{Aharonian2004} and \citet{Eungwanichayapant2009}, while EBL attenuation is implemented in the code following \citet{Barnard2024} with the EBL model by \citet{dominguez_extragalactic_2011}.

The Markov chain Monte Carlo (MCMC) optimisation technique is embedded in \textsc{naima} to find the best-fit model and associated best-fit parameters \citep{ForemanMackey2013}. The MCMC is implemented on the epoch with maximum data coverage. The function automatically takes some initial parameters as constant, such as $E_{k,\rm iso}$, $n_0$ (for ISM), $\dot{M}_w$ and $v_w$ (for wind), $z$, $t_{\rm start}$ and $t_{\rm stop}$. We consider five parameters as priors for the model fit such as the fraction of energy going into the electron and magnetic field log($\epsilon_e$) and log($\epsilon_B$), respectively, the cut-off energy log($E_c$), break energy log($E_b$) and power-law index $p$ of the electron distribution. The parameter $\epsilon_B$ was calculated using the formula $\epsilon_B = \frac{B^2}{16\pi\Gamma^2 n m_p c^2}$, and $p$ is calculated using $p\equiv\alpha_1=\alpha_2-1$. \textsc{naima} accounts for the maximum energy of electrons for synchrotron emission by using the effect of synchrotron burn-off, which is due to the balancing of acceleration and synchrotron losses of electrons in the $B$-field (in units of Gauss), 
given by Eq.~[18] from \citet{Aharonian2000} as
\begin{eqnarray} \label{eq:Eburnoff_lim}
    E_{c,\rm limit} & = \left(\frac{3}{2}\right)^{3/4}\sqrt{\frac{1}{e^3 B}} m_{\rm e}^2 c^4 \eta^{-1/2}, 
\end{eqnarray}
where $\eta$ is an efficiency parameter.

The prior range of $E_c$ is selected after calculating the value for GRB~160821B. For each MCMC fit, the number of parallel walkers, burn-in iterations, post-burn-in steps, and computational threads can be specified. Model selection is carried out using Bayesian inference, allowing the identification of the most probable model from a set of candidates based on the data.

\section{Results}
\label{sec3:results}
\begin{figure*}
    \centering
    \begin{minipage}[t]{0.48\textwidth}
        \centering
        \includegraphics[width=\linewidth]{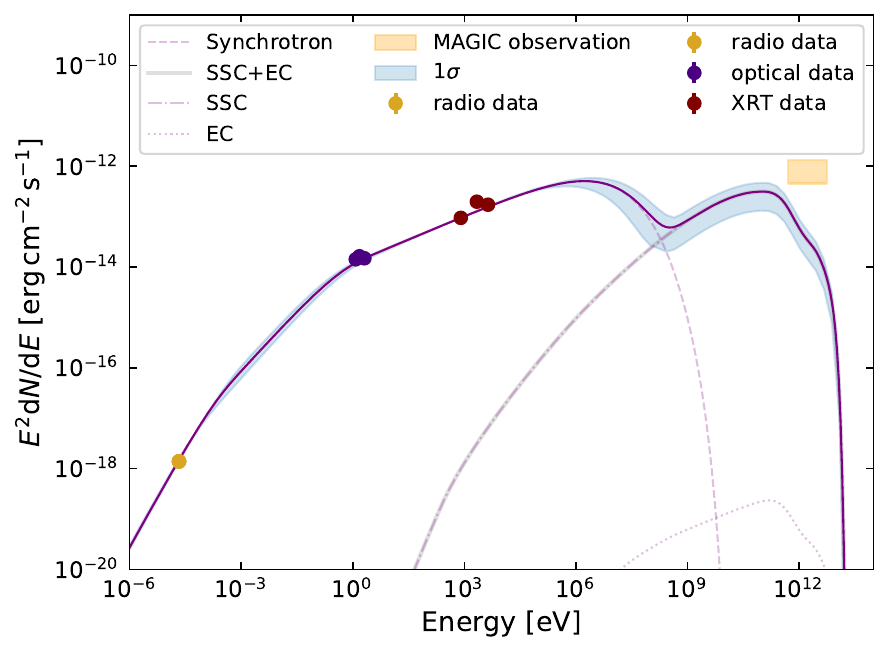}
        \caption{SED fitting of GRB~160821B in the ISM medium. The MCMC $1\sigma$-confidence band is shown over the model for the best-fit parameters mentioned in Table \ref{tab:model_parameters}. 
        }
        \label{fig:mcmc_160821b_ism}
    \end{minipage}
    \hfill
    \begin{minipage}[t]{0.48\textwidth}
        \centering
        \includegraphics[width=\linewidth]{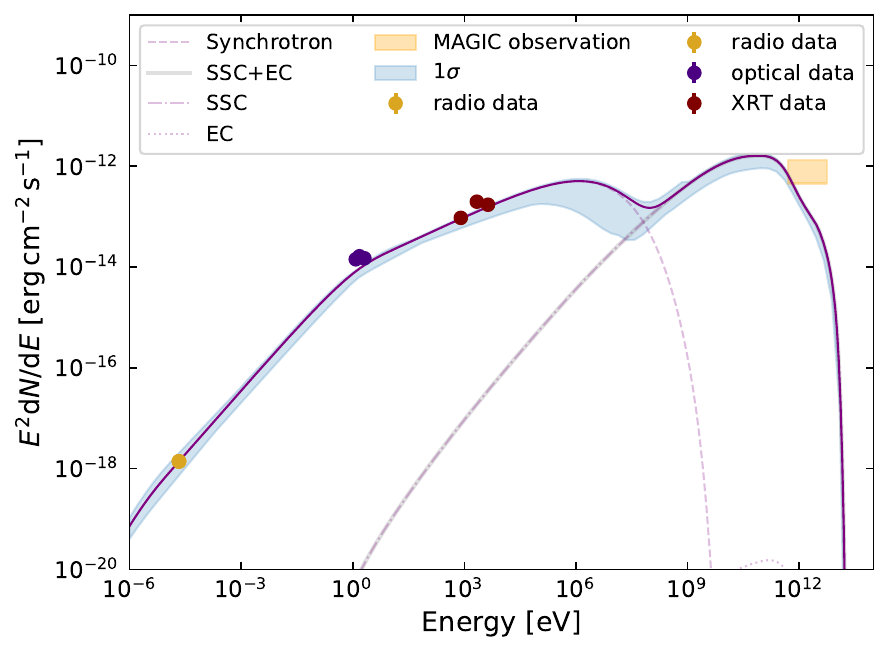}
        \caption{SED fitting of GRB~160821B for the wind medium. The MCMC $1\sigma$-confidence band is shown over the model for the best-fit parameters tabulated in Table \ref{tab:model_parameters}.}
        \label{fig:mcmc_160821b_wind}
    \end{minipage}
     
\end{figure*}

\begin{figure}
    \centering
    \begin{minipage}[t]{0.48\textwidth}
        \centering
        \includegraphics[width=\linewidth]{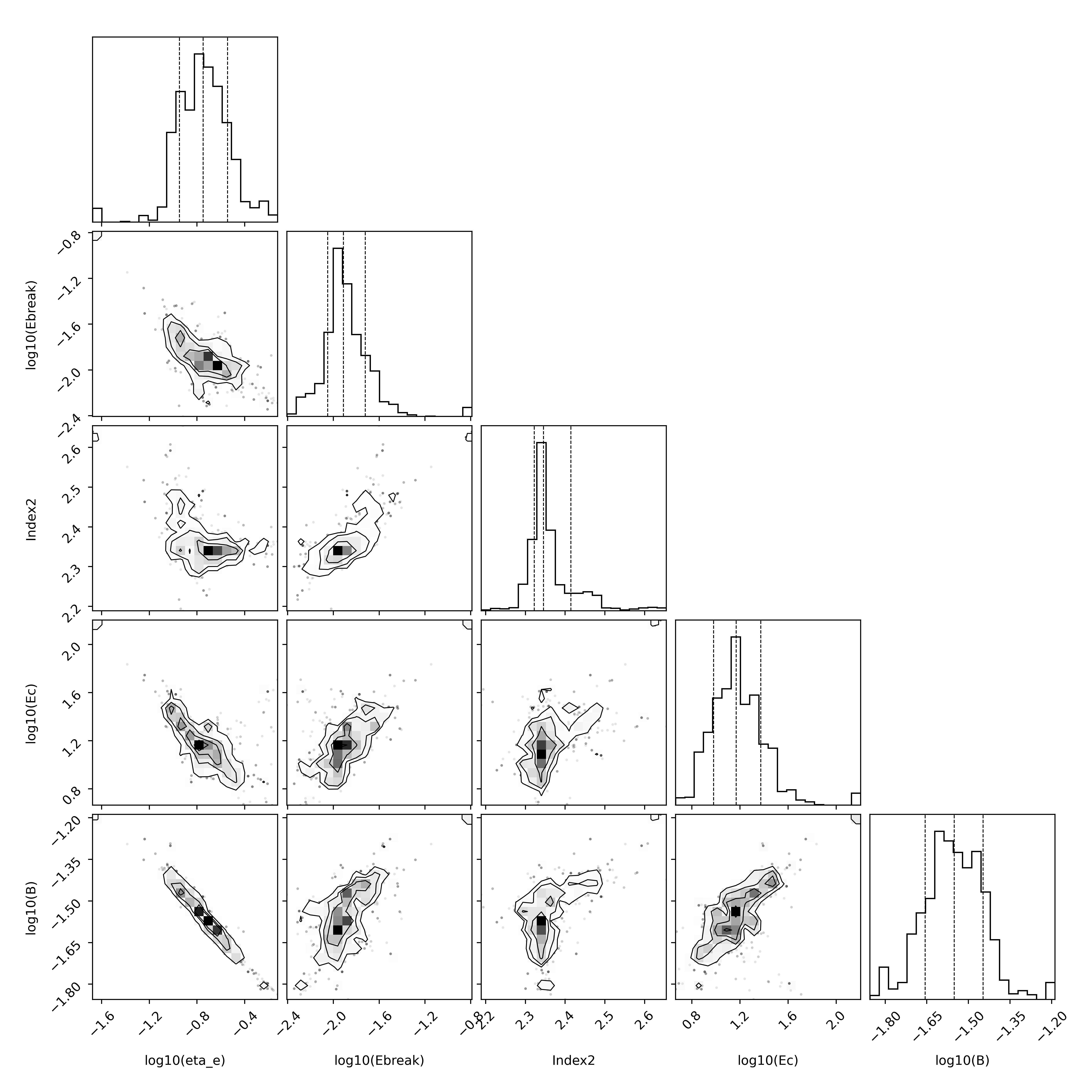}
        \caption{Corner plot for the ISM medium.}
        \label{fig:corner_160821b_ism}
    \end{minipage}
    \hspace{0.01\textwidth} 
    \begin{minipage}[t]{0.48\textwidth}
        \centering
        \includegraphics[width=\linewidth]{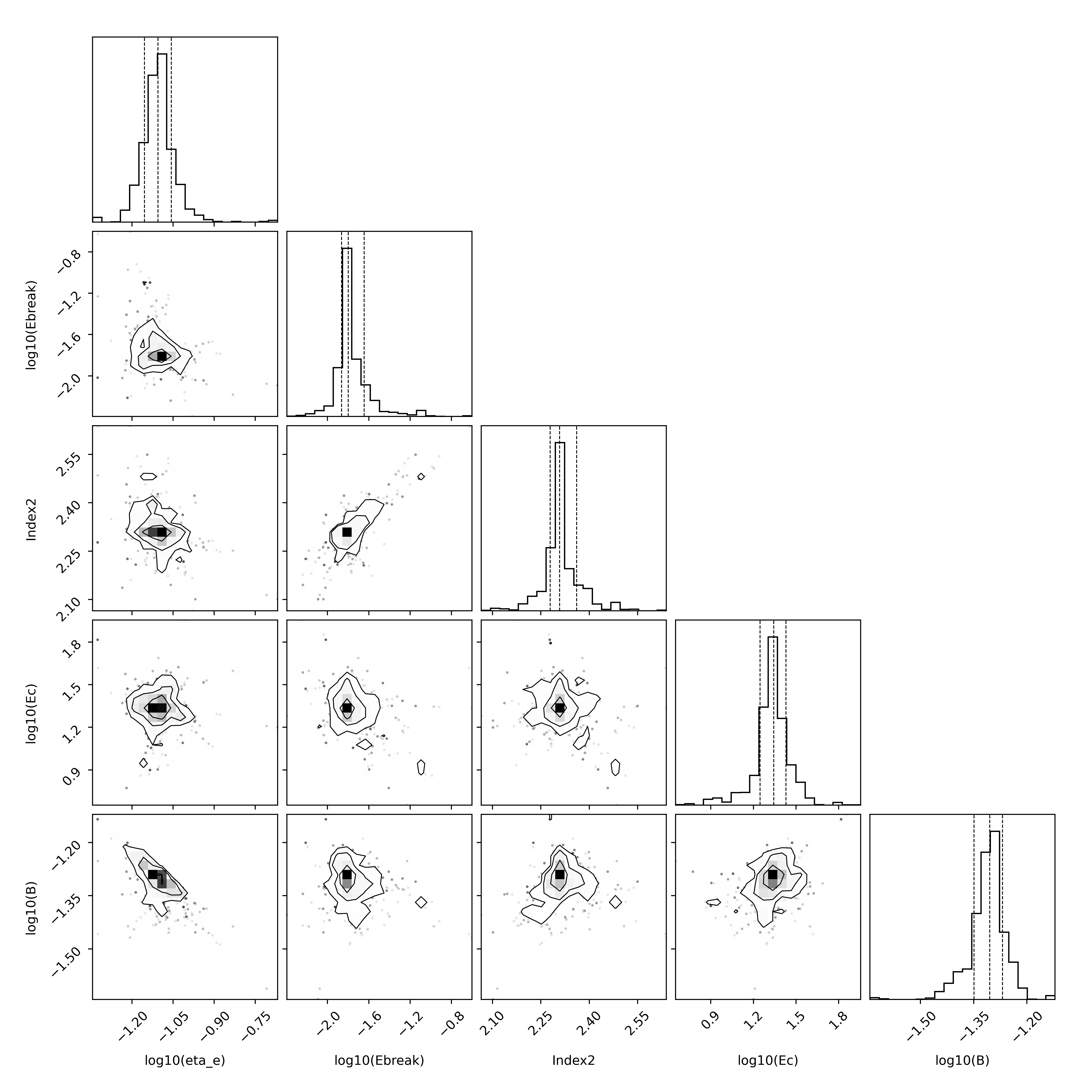}
        \caption{Corner plot for the wind medium.}
        \label{fig:corner_160821b_wind}
    \end{minipage}
    
\end{figure}

GRB~160821B has a rich dataset from across the electromagnetic spectrum. We obtained the almost simultaneous dataset from VHE to radio from \citet{Acciari2021ApJ..90A}. We perform a Bayesian information criterion (BIC) using MCMC for the dataset between 1.7 - 4 hours since the burst, to find the best fit model and obtain the corresponding best-fit parameters. The SEDs and corresponding corner plots are shown in Figs.~\ref{fig:mcmc_160821b_ism}, \ref{fig:corner_160821b_ism}, \ref{fig:mcmc_160821b_wind}, and \ref{fig:corner_160821b_wind}, for the ISM and wind medium, respectively. Although both the mediums are providing a good fit, the flux of the ISM model is slightly under predicting the MAGIC data, while the wind model successfully fits the VHE to radio data. So from the SED fitting, well constrained corner plots, and BIC values, we can clearly state that wind medium is fitting better for GRB~160821B.

As inferred from the afterglow model, the best fit parameter values for both the ISM and wind are very close to each other as mentioned in Table \ref{tab:model_parameters}. Both the mediums satisfy the lower isotropic energy with high ambient medium density. We calculated the value of $\Gamma_0$ = 197 and 273 for the ISM and wind medium, respectively.

\begin{table*}[ht]
\centering
\caption{Summary of the model parameter values that best fit the observations for each combination of GRB and density scenario. Bold values were held fixed during the fitting process, except for $\Gamma_0$ and $\epsilon_B$, which were calculated based on other parameters (see the main text for details). We assumed $E_{\rm k, iso}$ to be five times the isotropic-equivalent $\gamma$-ray energy release $E_{\gamma,\rm iso}$.}
\label{tab:model_parameters}
\resizebox{\textwidth}{!}{%
\begin{tabular}{l|ccccccccccccc|}
\hline \hline
Parameters & $E_{\mathrm{k,iso}}$ & $\boldsymbol{n_0}$  & $\dot{M}_{\rm w}$ & $\boldsymbol{v_{\rm w}}$ & $\boldsymbol{A_*}$ &$\Gamma_0$& log($\boldsymbol{\epsilon_B}$) & log($\epsilon_e$) & log($E_b$/TeV)  & $p$ & log($E_c$/TeV) & log($B$/G) \\
 & (erg) & (cm$^{-3}$) & (M$_\odot$/yr) & (cm/s) &  &  &  &  &  &  &    \\
\hline
ISM  & $4\times10^{50}$ & 10 & -- & -- & -- & 197.86 & $-4.40_{-0.25}^{+0.20}$ & $-0.95_{-0.08}^{+0.02}$ & $-1.82_{-0.09}^{+0.16}$ & $2.36_{-0.03}^{+0.07}$ & $1.35\pm0.15$ & $-1.45\pm0.08$  \\
Wind & $4\times10^{50}$ & -- & $2\times10^{-5}$ & $10^8$ & 2 & 273.01 & $-5.49_{-0.14}^{+0.04}$ & $-1.11\pm0.05$ & $-1.80_{-0.07}^{+0.15}$ & $2.31_{-0.03}^{+0.05}$ & $1.34\pm0.09$ & $-1.31\pm0.04$ \\
\hline
\end{tabular}%
}
\end{table*}

\section{Discussion \& conclusions}
\label{sec:conclusion}

\begin{figure}
    \centering
	\includegraphics[width=0.8\columnwidth]{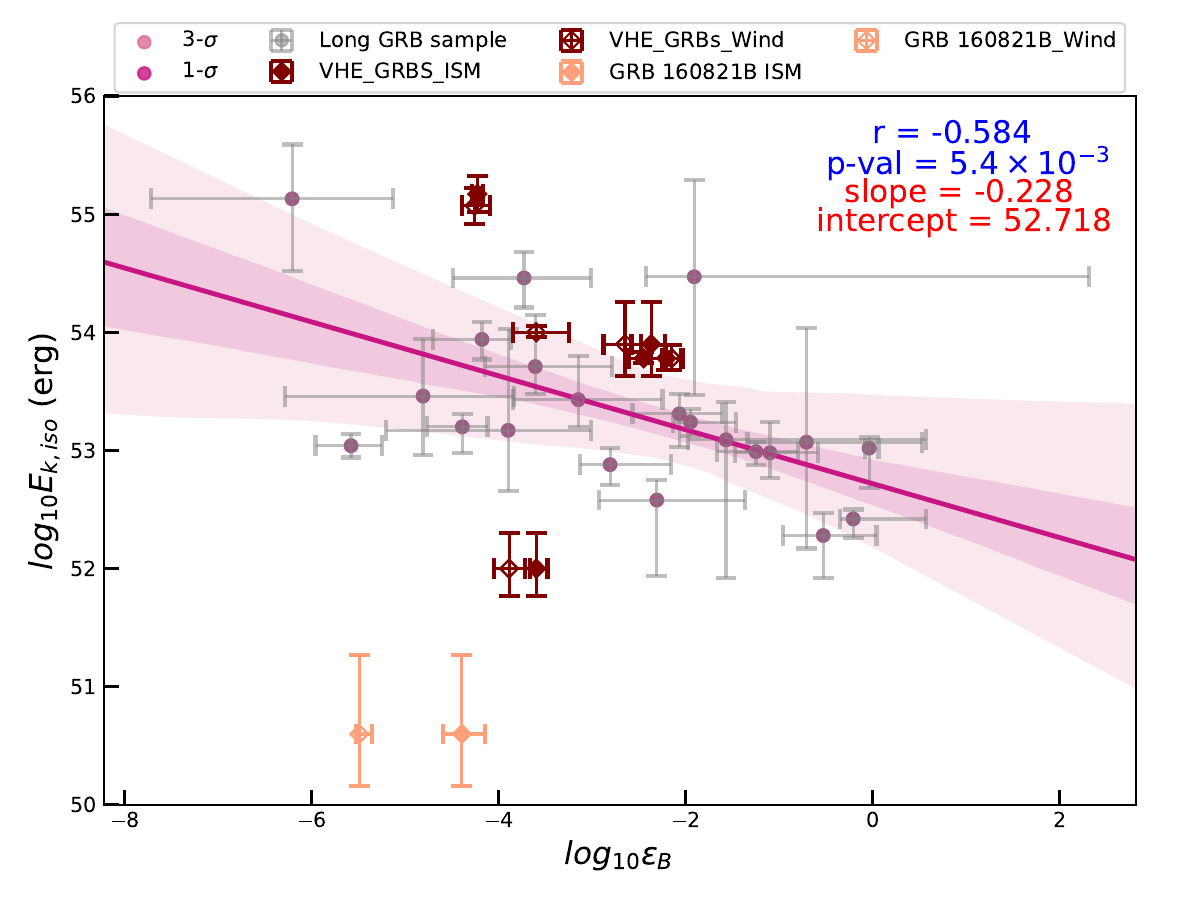}
    \caption{Correlation plot between $\epsilon_B$ and $E_{\rm k,iso}$, for the high energetic GRB sample. The diamonds and circles represent VHE detected long-durations and other high energetic long-durations mentioned in \citet{Cenko2011, Aksulu2022}. The shaded regions represent {$3\sigma$ and $1\sigma$ confidence interval respectively.} }
    \label{fig:corr}
\end{figure}

GRB~160821B, at a redshift $z = 0.161$, is the only short GRB for which VHE photon was detected. MAGIC telescope started following up on the field of GRB~160821B from 24 s to 4 hours after the burst, which led to 0.5 TeV photon detection at 3-$\sigma$ confidence level. Multi-wavelength SEDs were constructed from radio to VHE in between 1.7 - 4 hours since the burst and the data was collected from \citet{Acciari2021ApJ..90A}.

Since the first detection of VHE photons from GRBs, various models have been proposed to interpret their broadband afterglow emission. \citep{derishev2021ApJ.135D, Joshi2021, 2019Natur.575..455M} introduced a single zone model for GRB~190114C. A more complex multi-zone model incorporating the structured jet and off axis scenario was used by \citet{Salafia2022} for GRB~190829A. Eventually \citet{Ren2023} considered a structured jet morphology with core and wing configuration, and their transition to explain the complex evolution of GRB~221009A. But for GRB~160821B, \citet{Acciari2021ApJ..90A} used the analytical approach with single zone model with synchrotron and SSC, but they failed to explain the VHE photons with that approach. In contrast, \citet{2021ApJ...908L..36Z} demonstrated that synchrotron and EC model is explaining the VHE emission properly, though it lacked a robust optimisation framework for parameter estimation.

Our model adopts a single zone model for both the ISM and wind environments, considering synchrotron, SSC, and EC processes. We employed the MCMC optimisation technique to obtain the best fit model. Fig.~\ref{fig:mcmc_160821b_ism} and \ref{fig:mcmc_160821b_wind} clearly show that the contribution of EC is much lower than SSC, in contrast to the findings of \citet{2021ApJ...908L..36Z}. The wind model is preferred over ISM, as the ISM is clearly underestimating the VHE data as proposed in \citet{Acciari2021ApJ..90A}. In both scenarios the model parameters are comparable to each other, as mentioned in Table \ref{tab:model_parameters}. GRB~160821B falls under low energetic GRB category with high ambient medium density. Such a high value of $n_0$ is quite unusual for short GRBs and could reflect a distinct host environment, potentially linked to the VHE photon production. GRB~160821B has a lower $\epsilon_B$ value, for both the ISM and wind medium, than the other short GRBs mentioned in the study by \citet{2015ApJ...815..102F}, suggesting inefficient amplification of magnetic fields in the shocked region. These low $\epsilon_B$ values increase the SSC component's relative strength, providing a plausible explanation for the observed VHE emission in this event \citep{wang2010ssc, fraija2019ssc}. 

\citet{2015ApJ...815..102F} carried out multi-wavelength modelling of short GRBs over a decade (from 2005 to 2014), but they have chosen $\epsilon_B$ as a constant parameter. In the absence of any other comprehensive study of model parameter distributions for short GRBs, we use the log($E_{\rm k, iso}$) – log($\epsilon_B$) correlation presented by \citet{2025arXiv250518041B} for long GRBs to examine the location of GRB~160821B in that parameter space (see Fig.\ref{fig:corr}). As expected, the isotropic energy of GRB~160821B is much lower than the already detected long GRBs (with or without VHE emission). However, the $\epsilon_B$ value appears to be consistent with other VHE detected GRBs, which suggests a common trait between all the VHE GRBs. In conclusion, the single zone model with synchrotron and SSC for the wind model best describe the VHE emission for short GRB~160821B. Due to the lack of VHE detection from other short GRBs, it is hard to predict the radiation mechanism associated with the VHE photon detection in short GRBs. A statistically significant sample of VHE-detected short GRBs expected in the CTA era \citep{CTAConsortium:2017dvg} will be instrumental in constraining the underlying particle acceleration and radiation processes in these extreme transients. 
\\

\textit{Acknowledgment:} This research was supported by the National Research Foundation (NRF) of South Africa through a BRICS Multilateral Grant with number 150504 and an NRF Postdoctoral Fellowship to A.G.\ with grant number 2205056709.


\begin{thebibliography}{99}
\bibitem[Abdalla et al.(2019)]{Abdalla2019} Abdalla, H., Adam, R., Aharonian, F., et al.\ 2019, nat, 575, 7783, 464. doi:10.1038/s41586-019-1743-9.

\bibitem[H.~E.~S.~S. Collaboration et al.(2021)]{HESS:2021dbz} H.~E.~S.~S. Collaboration, Abdalla, H., Aharonian, F., et al.\ 2021, Science, 372, 6546, 1081. doi:10.1126/science.abe8560.

\bibitem[Abe et al.(2024)]{Abe2024} Abe, H., Abe, S., Acciari, V.~A., et al.\ 2024, mnras, 527, 3, 5856. doi:10.1093/mnras/stad2958.

\bibitem[Acciari et al.(2021)]{Acciari2021ApJ..90A} Acciari, V.~A., Ansoldi, S., Antonelli, L.~A., et al.\ 2021, apj, 908, 1, 90. doi:10.3847/1538-4357/abd249

\bibitem[Aharonian \& Atoyan(1981)]{1981ApnSS..79..321A} Aharonian, F.~A. \& Atoyan, A.~M.\ 1981, apss, 79, 2, 321. doi:10.1007/BF00649428

\bibitem[Aharonian et al.(2010)]{2010PhRvD..82d3002A} Aharonian, F.~A., Kelner, S.~R., \& Prosekin, A.~Y.\ 2010, prd, 82, 4, 043002. doi:10.1103/PhysRevD.82.043002



\bibitem[Cherenkov Telescope Array Consortium et al.(2019)]{CTAConsortium:2017dvg} Cherenkov Telescope Array Consortium, Acharya, B.~S., Agudo, I., et al.\ 2019, . doi:10.1142/10986

\bibitem[Aharonian(2000)]{Aharonian2000} Aharonian, F.~A.\ 2000, na, 5, 7, 377. doi:10.1016/S1384-1076(00)00039-7

\bibitem[Aharonian(2004)]{Aharonian2004} Aharonian, F.~A.\ 2004, . doi:10.1142/4657

\bibitem[Aksulu et al.(2022)]{Aksulu2022} Aksulu, M.~D., Wijers, R.~A.~M.~J., van Eerten, H.~J., et al.\ 2022, mnras, 511, 2, 2848. doi:10.1093/mnras/stac246

\bibitem[Banerjee et al.(2025)]{Banerjee2025} Banerjee, B., Macera, S., De Santis, A.~L., et al.\ 2025, aap, 701, A68. doi:10.1051/0004-6361/202554813

\bibitem[Barnard et al.(2024)]{Barnard2024} Barnard, M., Razzaque, S., \& Joshi, J.~C.\ 2024, mnras, 527, 4, 11893. doi:10.1093/mnras/stad3985

\bibitem[Barnard et al.(2025)]{2025arXiv250518041B} Barnard, M., Ghosh, A., Joshi, J.~C., et al.\ 2025, , arXiv:2505.18041. doi:10.48550/arXiv.2505.18041


\bibitem[Berti \& Carosi(2022)]{2022Galax..10...67B} Berti, A. \& Carosi, A.\ 2022, Galaxies, 10, 3, 67. doi:10.3390/galaxies10030067

\bibitem[Blanch et al.(2020)]{2020GCN.28659....1B} Blanch, O., Gaug, M., Noda, K., et al.\ 2020, GRB Coordinates Network, Circular Service, No. 28659, 28659, 1. 

\bibitem[Blandford \& McKee(1976)]{Blandford1976} Blandford, R.~D. \& McKee, C.~F.\ 1976, Physics of Fluids, 19, 1130. doi:10.1063/1.861619

\bibitem[Cao et al.(2023)]{doi:10.1126/sciadv.adj2778} Cao, Z., Aharonian, F., An, Q., et al.\ 2023, Science Advances, 9, 46, eadj2778. doi:10.1126/sciadv.adj2778


\bibitem[Cenko et al.(2010)]{Cenko2011} Cenko, S.~B., Frail, D.~A., Harrison, F.~A., et al.\ 2010, apj, 711, 2, 641. doi:10.1088/0004-637X/711/2/641

\bibitem[Derishev \& Piran(2021)]{derishev2021ApJ.135D} Derishev, E. \& Piran, T.\ 2021, apj, 923, 2, 135. doi:10.3847/1538-4357/ac2dec

\bibitem[Dermer et al.(2000)]{dermer_50GeV_IC} Dermer, C.~D., Chiang, J., \& Mitman, K.~E.\ 2000, apj, 537, 2, 785. doi:10.1086/309061

\bibitem[Dom{\'\i}nguez et al.(2011)]{dominguez_extragalactic_2011} Dom{\'\i}nguez, A., Primack, J.~R., Rosario, D.~J., et al.\ 2011, mnras, 410, 4, 2556. doi:10.1111/j.1365-2966.2010.17631.x


\bibitem[Eungwanichayapant \& Aharonian(2009)]{Eungwanichayapant2009} Eungwanichayapant, A. \& Aharonian, F.\ 2009, International Journal of Modern Physics D, 18, 6, 911. doi:10.1142/S0218271809014832

\bibitem[Fong et al.(2015)]{2015ApJ...815..102F} Fong, W., Berger, E., Margutti, R., et al.\ 2015, apj, 815, 2, 102. doi:10.1088/0004-637X/815/2/102

\bibitem[Foreman-Mackey et al.(2013)]{ForemanMackey2013} Foreman-Mackey, D., Hogg, D.~W., Lang, D., et al.\ 2013, pasp, 125, 925, 306. doi:10.1086/670067

\bibitem[Fraija et al.(2019)]{fraija2019ssc} Fraija, N., Barniol Duran, R., Dichiara, S., et al.\ 2019, apj, 883, 2, 162. doi:10.3847/1538-4357/ab3ec4

\bibitem[Gill \& Granot(2022)]{2022Galax..10...74G} Gill, R. \& Granot, J.\ 2022, Galaxies, 10, 3, 74. doi:10.3390/galaxies10030074

\bibitem[Isravel et al.(2023)]{2023ApJ...955...70I} Isravel, H., Pe'er, A., \& B{\'e}gu{\'e}, D.\ 2023, apj, 955, 1, 70. doi:10.3847/1538-4357/acec73


\bibitem[Joshi \& Razzaque(2021)]{Joshi2021} Joshi, J.~C. \& Razzaque, S.\ 2021, mnras, 505, 2, 1718. doi:10.1093/mnras/stab1329

\bibitem[Khangulyan et al.(2014)]{2014ApJ...783..100K} Khangulyan, D., Aharonian, F.~A., \& Kelner, S.~R.\ 2014, apj, 783, 2, 100. doi:10.1088/0004-637X/783/2/100


\bibitem[Klinger et al.(2024)]{Klinger_2024} Klinger, M., Yuan, C., Taylor, A.~M., et al.\ 2024, apj, 977, 2, 242. doi:10.3847/1538-4357/ad9392

\bibitem[MAGIC Collaboration et al.(2019)]{2019Natur.575..455M} MAGIC Collaboration, Acciari, V.~A., Ansoldi, S., et al.\ 2019, nat, 575, 7783, 455. doi:10.1038/s41586-019-1750-x

\bibitem[Miceli \& Nava(2022)]{2022Galax..10...66M} Miceli, D. \& Nava, L.\ 2022, Galaxies, 10, 3, 66. doi:10.3390/galaxies10030066

\bibitem[Misra et al.(2021)]{Misra2021MNRAS_5685M} Misra, K., Resmi, L., Kann, D.~A., et al.\ 2021, mnras, 504, 4, 5685. doi:10.1093/mnras/stab1050

\bibitem[Murase et al.(2010)]{2010MNRAS.402L..54M} Murase, K., Toma, K., Yamazaki, R., et al.\ 2010, mnras, 402, 1, L54. doi:10.1111/j.1745-3933.2009.00799.x

\bibitem[Nakar et al.(2009)]{Nakar:2009er} Nakar, E., Ando, S., \& Sari, R.\ 2009, apj, 703, 1, 675. doi:10.1088/0004-637X/703/1/675

\bibitem[Panaitescu \& Kumar(2000)]{Panai2000ApJ543} Panaitescu, A. \& Kumar, P.\ 2000, apj, 543, 1, 66. doi:10.1086/317090

\bibitem[Ren et al.(2023)]{Ren2023} Ren, J., Wang, Y., Zhang, L.-L., et al.\ 2023, apj, 947, 2, 53. doi:10.3847/1538-4357/acc57d

\bibitem[Sahu \& Fort{\'\i}n(2020)]{sahu2020ApJ.41S} Sahu, S. \& Fort{\'\i}n, C.~E.~L.\ 2020, apjl, 895, 2, L41. doi:10.3847/2041-8213/ab93da

\bibitem[Sahu et al.(2022)]{2022ApJ...929...70S} Sahu, S., Valadez Polanco, I.~A., \& Rajpoot, S.\ 2022, apj, 929, 1, 70. doi:10.3847/1538-4357/ac5cc6

\bibitem[Salafia et al.(2022)]{Salafia2022} Salafia, O.~S., Ravasio, M.~E., Yang, J., et al.\ 2022, apjl, 931, 2, L19. doi:10.3847/2041-8213/ac6c28

\bibitem[Sari \& Esin(2001)]{sari_IC_paper} Sari, R. \& Esin, A.~A.\ 2001, apj, 548, 2, 787. doi:10.1086/319003

\bibitem[Wang et al.(2010)]{wang2010ssc} Wang, X.-Y., He, H.-N., Li, Z., et al.\ 2010, apj, 712, 2, 1232. doi:10.1088/0004-637X/712/2/1232

\bibitem[Zabalza(2015)]{2015ICRC...34..922Z} Zabalza, V.\ 2015, 34th International Cosmic Ray Conference (ICRC2015), 34, 922. doi:10.22323/1.236.0922

\bibitem[Zhang \& M{\'e}sz{\'a}ros(2001)]{2001ApJ...559..110Z} Zhang, B. \& M{\'e}sz{\'a}ros, P.\ 2001, apj, 559, 1, 110. doi:10.1086/322400

\bibitem[Zhang et al.(2021)]{2021ApJ...908L..36Z} Zhang, B.~T., Murase, K., Yuan, C., et al.\ 2021, apjl, 908, 2, L36. doi:10.3847/2041-8213/abe0b0

\bibitem[Zhang et al.(2021)]{zhang2021ApJ..55Z} Zhang, B.~T., Murase, K., Veres, P., et al.\ 2021, apj, 920, 1, 55. doi:10.3847/1538-4357/ac0cfc

\bibitem[Zhang et al.(2023)]{2023ApJ...947L..14Z} Zhang, B.~T., Murase, K., Ioka, K., et al.\ 2023, apjl, 947, 1, L14. doi:10.3847/2041-8213/acc79f






\end{thebibliography}
\end{document}